\documentclass[12pt]{article}

\usepackage[cp1251]{inputenc}
\usepackage{amsfonts}
\usepackage{amssymb}
\usepackage{amsthm}
\usepackage{amsmath}
\usepackage{newlfont}
\usepackage{graphicx}

\date{}
\title{Form factors of tetraquarks}
\author{ S. M. Gerasyuta$^{1,2}$, M. A. Durnev$^{1}$ \\ {\em $^{1}$  Department of Theoretical Physics}\\ 
\em {St. Petersburg State University, 198904,}\\ {\em St. Petersburg, Russia}\\
{\em $^{2}$ Department of Physics, LTA, 194021,} \\
\em {St. Petersburg, Russia}\\
} 
\scrollmode

\begin{document}
\maketitle

\begin{abstract}
{\small 
\noindent     The electromagnetic form factors of tetraquarks are calculated in the framework
of relativistic quark model at small and intermediate momentum transfers $Q^2 \le$ 1 GeV$^2$. 
The charge radii of $X(3872)$ and $X(3940)$ tetraquarks are determined.  \\ 
}
\end{abstract}
{\bf 1. Introduction \\}

The interest in the heavy quark spectroscopy has been revived due to the discovery of new heavy 
hadrons, containing the charmed quarks, over the last few years [1, 2]. In 2003 a new resonance named
$X$(3872) was reported by the Belle Collaboration in the invariant mass distribution of
$J/\psi\pi^{+}\pi^{-}$ mesons produced in $B^{\pm}\to K^{\pm}X(3872)\to K^{\pm}J/\psi\pi^{+}\pi^{-}$ 
decays. It appeared as a narrow peak with a mass 3871.2 $\pm$ 0.5 MeV and a width $\Gamma<$  2.3 
MeV [3]. This state was confirmed by BaBar [4], CDF [5] and D0 Collaborations [6]. 
%

In July 2005 Belle claimed the observation of a new charmonium resonance named $X(3940)$
with a mass of 3943 $\pm$ 6 $\pm$ 6 MeV and a total width of less than 52 MeV [7]. 

In the early 80's, Gelmini [8] studied the $S$-wave $c\bar c n\bar n$ states using the one-gluon-exchange
potential and the virtual annihilation of color pairs, obtaining some candidates that could lie below any of 
the dissociation channels. Chao [9] explored the decay, hadronic production, production $e^{+}e^{-}$
annihilation, and photoproduction of various types of $c\bar c n\bar n$ states, using the quark-gluon
model proposed by Chan and Hogaasen [10]. 
After the discovery of the $X(3872)$ the question about the possible existence of
$c\bar c n\bar n$ bound states was posed again. Maiani {\em et al.} [11] constructed a model of the
$X(3872)$ in terms of diquark-antidiquark degrees of freedom. Using the $X(3872)$ as input they predict
other $c\bar c n\bar n$ states with quantum numbers $0^{++}, 1^{+-}$ and $2^{++}$. Ebert {\em et al.}
[12] addressed heavy tetraquarks with hidden charm in a diquark-antidiquark relativistic quark model,
concluding that the $X(3872)$ could be identified with the $1^{++}$ neutral charmed tetraquark.

The consideration of relativistic effects in the composite systems is sufficiently
important when the quark structure of the hadrons is studied [13-21].
The dynamical variables (form factors, scattering amplitudes) of composite
particles can be expressed in terms of the Bethe-Salpeter equations or
quasipotentials. The form factors of the composite particles were considered
by a number of authors, who have in particular applied a ladder approximation
for the Bethe-Salpeter equation [22], ideas of conformal invariance
[23], a number of results was obtained in the framework of three-dimensional
formalisms [24]. It seems that an application of the dispersion integrals over
the masses of the composite particles may be sufficiently convenient to the
description of the relativistic effects in the composite systems. On the one
hand, the dispersion relation technique is relativistically invariant one and
it is not determined with a consideration of any specific coordinate system.
On the other hand, there is no problem of additional states arising,
because contributions of intermediate states are controlled in the dispersion
relations. The dispersion relation technique allows to determine the form
factors of the composite particles [25].

     In papers [26, 27] the nucleon form factors were calculated in the dispersion relation technique.
     In paper [26] the relativistic generalization of the Faddeev equations was constructed
in the form of dispersion relations in the pair energy of two interacting particles
and the integral equations were obtained for the three-particle amplitudes
of  $S$-wave baryons: for the octet $J^P=\dfrac {1} {2} ^+$ 
and the decuplet $J^P=\dfrac {3} {2} ^+$. 
The approximate solution of the relativistic three-particle problem using the
method based on the extraction of the leading singularities of the scattering
amplitudes about  $s_{ik}=4m^2$ was proposed. The three-quark amplitudes
given in Refs. [26, 28] could be used for the calculation of electromagnetic
nucleon form factors at small and intermediate momentum transfers [27].
Using these results the proposed approach was generalized to
the case of five particles in the recent paper [29].    

In the present paper the computational scheme of the electromagnetic
form factors of the tetraquarks, consisted of four particles, in the
infinite momentum frame is given.

Section II is devoted to the calculation of electromagnetic 
form factors of $N$ particle system and a special case $N=4$ in the infinite momentum frame. 
The calculation results of electric form factors of the tetraquark states $X(3872)$ and $X(3940)$ 
are given in Section III. The last section is devoted to our discussion and conclusion. \\ \\

\newpage 
{\bf 2. The calculation of electromagnetic form factors of tetraquarks
in the infinite momentum frame \\}

The approach used in [29] for considering of form factors of exotic baryons with isospin
$I=5/2$ and based on the transition from Feynman amplitude to the dispersion integration
over the masses of the composite particles, may be extended for multiquark systems 
containing $N$ quarks.

Let us consider the electric form factor of a system of $N$ particles, 
shown in Fig.1{\em a}. The momentum of the system is treated to be large:
$P_{z} \to \infty$, the momenta $P=k_{1}+k_{2}+...+k_{N}$ and $P'=P+q$ correspond to 
the initial and final momenta of the system. Let us assume
$P=(P_{0}, \mathbf P_{\perp}=0, P_{z})$  and $P'=(P'_{0}, \mathbf P'_{\perp}, P'_{z})$, 
where $P^2=s,\quad P'^2=s'$. 
Then we have some conservation laws for the input momenta:\\[-30pt]
 {\multlinegap=0pt \begin{multline} \sum\limits_{i=1}^{N}\mathbf k_{i\perp}=0, \\
     \shoveleft{P_{z}-\sum\limits_{i=1}^{N}k_{iz}=P_{z}(1-\sum\limits_{i=1}^{N}x_{i})=0,} \\
     \shoveleft{P_{0}-\sum\limits_{i=1}^{N}k_{i0}=P_{z}(1-\sum\limits_{i=1}^{N}x_{i}) +
    \dfrac {1} {2P_{z}}\left.\left(s-\sum\limits_{i=1}^{N}\dfrac {m^2_{i\perp}} {x_{i}} 
    \right) \right. =0,} \\
    \shoveleft{m^2_{i\perp}=m^2+\mathbf k^2_{i\perp}, \quad x_{i}=\dfrac {k_{iz}} {P_{z}}, \quad i=1,2,...,N. }
     \qquad \qquad \qquad \qquad \qquad  \end{multline} } \\[-30pt]
By analogy for the output momenta :\\[-30pt]
{\multlinegap=0pt \begin{multline}  
    \mathbf k'_{1\perp}+\sum\limits_{i=2}^{N}\mathbf k_{i\perp}-
    \mathbf q_{\perp}=0, \\ 
    \shoveleft{P'_{z}-k'_{1z}-\sum\limits_{i=2}^{N} k_{iz}=P_{z}(z-x'_{1}-\sum\limits_{i=2}^{N} x_{i})=0,} \\
    \shoveleft{P'_{0}-k'_{10}-\sum\limits_{i=2}^{N} k_{i0}=P_{z}(z-x'_{1}-\sum\limits_{i=2}^{N} x_{i}) +
    \dfrac {1} {2P_{z}} \left.\left(\dfrac {s'+\mathbf q^2_{\perp}} {z}-\dfrac {m'^2_{1\perp}} {x'_{1}} -  
    \sum\limits_{i=2}^{N}\dfrac {m^2_{i\perp}} {x_{i}} \right) \right. =0,} \\
    \shoveleft{x'_{1}=\dfrac {k'_{1z}} {P_{z}}, \quad m'^2_{1\perp}=m^2_{1}+\mathbf k'^2_{1\perp}. }
\qquad \qquad \qquad \qquad \qquad \qquad \qquad \qquad \qquad \end{multline} }   

It is introduced in (1) and (2) $\mathbf q_{\perp} \equiv \mathbf P'_{\perp}$ and $z=\dfrac {P'_{z}} 
{P_{z}}=\dfrac {s'+s-q^2} {2s}$. The form factor of the $N$-quark system can be obtained with the help of the
double dispersion integral [25]: 
$$  F(q^2)= \int\limits_{(\sum\limits_{i=1}^{N} m_{i})^2}^{\Lambda_{s}} \dfrac {ds\; ds'} {4\pi^2} 
\dfrac {disc_{s} disc_{s'} F(s,s',q^2)} {(s-M^2)(s'-M^2)}, \qquad \qquad \qquad \qquad \qquad \eqno (3) $$
$$ disc_{s} disc_{s'} F(s,s',q^2)=GG'\int d\rho (P,P',k_{1},k_{2},...,k_{N-1}) \qquad \qquad \qquad \eqno (4) $$
The invariant phase space $d\rho (P,P',k_{1},k_{2},...,k_{N-1})$ , which enters in the double
dispersion integral, has the form [25]:
 {\multlinegap=0pt \begin{multline}
d\rho (P,P',k_{1},k_{2},...,k_{N-1})=d\Phi^{(N)}(P,k_{1},k_{2},...,k_{N})\times d\Phi^{(N)}
(P', k'_{1}, k'_{2},..., k'_{N})\times \\ \shoveleft{ \times \prod\limits_{l=2}^{N} (2\pi)^{3} 2k_{l0} 
\delta^3(\mathbf k_{l}-\mathbf k'_{l}),}
\qquad \qquad \qquad \qquad \qquad \qquad \qquad  \qquad \qquad \tag{5} \end{multline} }  
where the $N$-particle phase space is introduced:
$$d\Phi^{(N)}(P,k_{1},k_{2},...,k_{N})=(2\pi)^4\delta^4 \bigl(P-\sum\limits_{i=1}^{N} k_{i} \bigr)
\prod_{l=1}^{N} \dfrac{d^3 k_{l}} {(2\pi)^{3}2(k_{l0})^2}. \qquad \qquad \quad $$
After the transformation we have:
 {\multlinegap=0pt \begin{multline}
d\rho (P,P',k_{1},k_{2},...,k_{N-1})= \dfrac{1} {2^{N-1}(2\pi)^{3N-5}} \prod\limits_{l=1}^{N-1} \dfrac {dx_{l}} {x_{l}} 
d\mathbf k_{l\perp}  \times \dfrac {1} {z-1+x_{1}} \times 
\\ \shoveleft{\times \dfrac {1} {\bigl(1-\sum\limits_{i=1}^{N-1}x_{i}\bigr)} \times \delta 
\left.\left(s-\sum\limits_{i=1}^{N-1} \dfrac {m^2_{i\perp}} {x_{i}} -\dfrac {m^2_{N\perp}} 
{1-\sum\limits_{i=1}^{N-1} x_{i}} \right) 
\right. \times}  \\ \shoveleft{\times \delta \left.\left(\dfrac {s'+\mathbf q^2_{\perp}} {z} -\dfrac 
{m'^2_{1\perp}} {z-1+x_{1}} -\sum\limits_{i=2}^{N-1} \dfrac {m^2_{i\perp}} {x_{i}}-\dfrac {m^2_{N\perp}} 
{1-\sum\limits_{i=1}^{N-1} x_{i}}\right). \right.  }
\qquad \qquad \qquad \qquad  \raisetag{60pt} \tag{6} \end{multline} } \\
For the diquark-spectator (Fig.1{\em b}) the invariant phase space takes the following form similar to (6):
 {\multlinegap=0pt \begin{multline}
d\rho (P,P',k_{1},k_{2},...,k_{N-2})= \dfrac{\dfrac {1} {2} I_{N-1,N}} {2^{N-1}(2\pi)^{3N-5}} 
\prod\limits_{l=1}^{N-2} \dfrac {dx_{l}} {x_{l}} 
d\mathbf k_{l\perp}  \times \dfrac {1} {z-1+x_{1}} \times  \quad 
\\ \shoveleft{\times \dfrac {1} {\bigl(1-\sum\limits_{i=1}^{N-2}x_{i}\bigr)} \times \delta 
\left.\left(s-\sum\limits_{i=1}^{N-2} \dfrac {m^2_{i\perp}} {x_{i}} -\dfrac {m^2_{N-1,N\perp}} 
{1-\sum\limits_{i=1}^{N-2} x_{i}} \right) 
\right. \times}  \\ \shoveleft{\times \delta \left.\left(\dfrac {s'+\mathbf q^2_{\perp}} {z} -\dfrac 
{m'^2_{1\perp}} {z-1+x_{1}} -\sum\limits_{i=2}^{N-2} \dfrac {m^2_{i\perp}} {x_{i}}-\dfrac {m^2_{N-1,N\perp}} 
{1-\sum\limits_{i=1}^{N-2} x_{i}}\right), \right.}  
\qquad \qquad \qquad \qquad  \raisetag{60pt} \tag{7} \end{multline} } \\
where the phase space of the diquark is determined by $I_{N-1,N}$. In such a way the expressions for
any number $n$ ($n \le [\frac{N-1} {2}]$) of diquark-spectators may be obtained.
 
Let us consider the latest $\delta$-function in (6). Assuming that:\\[5pt]
$\sum\limits_{i=2}^{N-1} \dfrac {m^2_{i\perp}} {x_{i}} - \dfrac {m^2_{N\perp}} {1-\sum\limits_{i=1}^{N-1}x_{i}} =
s-\dfrac{m^2_{1\perp}} {x_{1}},$\\[5pt]
it is obvious that:\\
$\delta \left.\left(\dfrac {s'+\mathbf q^2_{\perp}} {z} -\dfrac 
{m'^2_{1\perp}} {z-1+x_{1}} -\sum\limits_{i=2}^{N-1} \dfrac {m^2_{i\perp}} {x_{i}}-\dfrac {m^2_{N\perp}} 
{1-\sum\limits_{i=1}^{N-1} x_{i}}\right) \right.  = \\
= \; \delta \left.\left(\dfrac {s'+\mathbf q^2_{\perp}} {z} -\dfrac 
{m'^2_{1\perp}} {z-1+x_{1}}-s+\dfrac{m^2_{1\perp}} {x_{1}} \right). \right. $\\[5pt] 
So we get an expression, which doesn't depend on $N$.
Hence the integration over $s'$ is carried out exactly in the same way as in the previous paper [29].
The expressions for $s$ and $\tilde s$, which are obviously depend on the number of particles $N$,
have the following form:\\[-40pt]
{\multlinegap=0pt \begin{multline}
s=\sum\limits_{i=1}^{N-2} \dfrac {m^2_{i\perp}} {x_{i}} + \dfrac {m^2_{N-1,N}+
\sum\limits_{i=1}^{N-2}k^2_{i\perp}+2\sum\limits_{i=1}^{j-1}\sum\limits_{j=2}^{N-2} 
\sqrt{k^2_{i\perp}k^2_{j\perp}}\cos{(\phi_{j}-\phi_{i})}} {1-\sum\limits_{i=1}^{N-2}x_{i}}, 
\\ \shoveleft{\tilde s=\sum\limits_{i=1}^{N-1} \dfrac {m^2_{i\perp}} {x_{i}} + \dfrac {m^2_{N}+
\sum\limits_{i=1}^{N-1}k^2_{i\perp}+2\sum\limits_{i=1}^{j-1}\sum\limits_{j=2}^{N-1} 
\sqrt{k^2_{i\perp}k^2_{j\perp}}\cos{(\phi_{j}-\phi_{i})}} {1-\sum\limits_{i=1}^{N-1}x_{i}}.}
\qquad \quad \tag{8} \end{multline} } \\[-20pt]
Having integrated the $\delta$-functions, we obtain the following expression for $N$-particle contribution
to the form factor (3):\\[-30pt]
{\multlinegap=0pt \begin{multline}
F_{0}=\dfrac {2} {2^{2(N-1)}(2\pi)^{3(N-1)}}\int\limits_{0}^{\Lambda_{k_{\perp}}}\; \prod\limits_{i=1}^{N-1}
dk^2_{i\perp} \; \int\limits_{0}^{1} \; \prod\limits_{i=1}^{N-1} dx_{i} \; \int\limits_{0}^{2\pi} \;
\prod\limits_{i=1}^{N-1} d\phi_{i} \; \times \\[-5pt] 
\shoveleft{\times \dfrac {1} { 
\prod\limits_{i=1}^{N-1}x_{i}(1-x_{i})} \;
 \dfrac {\tilde b \tilde \lambda + 1} {\tilde b + \tilde \lambda \tilde f}\;
 \dfrac {\theta(\Lambda_{s}-\tilde s) \theta(\Lambda_{s}-\tilde s')} {(\tilde s-M^2)(\tilde s'-M^2)},}
\qquad \qquad \qquad \qquad \qquad \tag{9} \end{multline} } \\[-20pt]
the following one is for the case with one diquark-spectator:\\[-30pt]
{\multlinegap=0pt \begin{multline}
F_{1}=\dfrac {2 I_{N-1,N}} {2^{2(N-1)}(2\pi)^{3(N-1)}}\int\limits_{0}^{\Lambda_{k_{\perp}}}\; \prod\limits_{i=1}^{N-2}
dk^2_{i\perp} \; \int\limits_{0}^{1} \; \prod\limits_{i=1}^{N-2} dx_{i} \; \int\limits_{0}^{2\pi} \;
\prod\limits_{i=1}^{N-2} d\phi_{i} \; \times \\[-5pt]
\shoveleft{\times \; \dfrac {1} {
\prod\limits_{i=1}^{N-2}x_{i}(1- x_{i}) } 
\dfrac { b \lambda + 1} { b +  \lambda f}\; 
 \dfrac {\theta(\Lambda_{s}-s) \theta(\Lambda_{s}- s')} {(s-M^2)(s'-M^2)} } 
\qquad \qquad \qquad \qquad \qquad \tag{10} \end{multline} } \\[-40pt]
{\multlinegap=0pt \begin{multline}
b=x_{1}+ \dfrac {m^2_{1\perp}} {sx_{1}}, \quad f=b^2-\dfrac {4k^2_{1\perp}\cos^2(\phi_{1})} {s},
\\[-0pt] \shoveleft{\lambda= \dfrac {-b+\sqrt{(b^2-f)\Bigl(1-\Bigl(\dfrac {s} {q^2}\Bigr)f \Bigr) }} {f}, \quad \; 
 s'=s+q^2(1+2\lambda);} \qquad \qquad \tag{11} \end{multline} } \\[-40pt]
{\multlinegap=0pt \begin{multline}
\tilde b=x_{1}+ \dfrac {m^2_{1\perp}} {\tilde sx_{1}}, \quad \tilde f=\tilde b^2-\dfrac {4k^2_{1\perp}
\cos^2(\phi_{1})} {\tilde s}, 
\\[-0pt] \shoveleft{\tilde \lambda= \dfrac{-\tilde b+\sqrt{(\tilde b^2-\tilde f)\Bigl 
(1-\Bigl (\dfrac {\tilde s} {q^2}\Bigr ) \tilde f \Bigr )}} {\tilde f}, \quad
\tilde s'=\tilde s+q^2(1+2\tilde \lambda).}  \qquad  \tag{12} \end{multline} } \\[-20pt]

     To find the tetraquark form factor one needs to put $N=4$ and to account the interaction
of each light quark with the external electromagnetic field using the form factor
of nonstrange quarks $f_{q}(q^2)$ [30]. For the electromagnetic form factor of tetraquark in the 
case of the normalization $G^{E}(0)=1$ we obtain:
$$G^{E}(q^2)=\dfrac {F^{E}(q^2)} {F^{E}(0)}=\dfrac {f_{q}(q^2)} {f_{q}(0)} \dfrac {J_{6}(q^2)+J_{9}
(q^2)} {J_{6}(0)+J_{9}(0)}, \qquad \qquad   \qquad   \qquad   \qquad  \quad  \eqno (13) $$
where:\\[-40pt]
{\multlinegap=0pt \begin{multline}
 J_{6}(q^2)=I_{34}\int\limits_{0}^{\Lambda_{k_{\perp}}} \prod\limits_{i=1}^{2} 
dk^2_{i\perp} \int\limits_{0}^{1} \prod \limits_{i=1}^{2} dx_{i} \int_{0}^{2\pi} \prod \limits_{i=1}^{2} 
d\phi_{i}\dfrac {1} {x_{1}(1-x_{1})x_{2}(1-x_{2})}\; \times 
\\[-0pt] \shoveleft{ \times \dfrac {b\lambda+1} {b+\lambda f} A^2_{3} \dfrac {\theta(\Lambda_{s}-s) \theta
(\Lambda_{s}-s')} {(s-M^2)(s'-M^2)},}
\\[-0pt] \shoveleft{
J_{9}(q^2)=\int\limits_{0}^{\Lambda_{k_{\perp}}} \prod \limits_{i=1}^{3} dk^2_{i\perp} \int
\limits_{0}^{1} \prod \limits_{i=1}^{3} dx_{i} \int_{0}^{2\pi} \prod \limits_{i=1}^{3} d\phi_{i}
\dfrac {1} {x_{1}(1-x_{1})x_{2}(1-x_{2})x_{3}(1-x_{3})} \times } 
\\[-0pt] \shoveleft{ \times \dfrac 
{\tilde b \tilde \lambda+1} {\tilde b+\tilde \lambda \tilde f} A^2_{1} \dfrac {\theta(\Lambda_{s}-
\tilde s)\theta(\Lambda_{s}-\tilde s')} {(\tilde s-M^2)(\tilde s'-M^2)}.} 
\qquad \qquad \qquad \qquad \qquad \qquad \qquad \qquad   \tag{14} \end{multline} } \\[-20pt]
$A_{n}$ ($n$ = 1,3) determine relative contributions of the subamplitudes 
in the total amplitude of the tetraquark [31].  \\

{\bf 3. Calculation results }\\

    The electric form factor of tetraquark is the sum of two terms (13). 
The phase space of the spectator contributes to the first term $I_{34}=9.11$ GeV$^2$. 
The vertex functions $G$ and $G'$ are taken in the middle point of the physical region. 
The mass of the quark $u$ is equal to $m_{u}=0.385$ GeV, and the mass of the $c$ quark is 
$m_{c}=1.586$ GeV. The cutoff parameter over the pair energy $\lambda=10$ is obtained in [31]. 
It is possible to calculate the dimensional  cutoff parameters over the total energy and the 
transvers momentum $\Lambda_{s} = 75$ GeV$^2$,\quad $\Lambda_{k^2_{i\perp}}=1.5m^2_{i}$ 
($m_{i}=m_{u}, \, m_{c}$) respectively. It is necessary to account that $\gamma$-quantum interacts only 
with the light dressed quark, which has own form factor [30]: for $u$  quark 
$f_{q}(q^2)=$exp$(\alpha_{q}q^2), \, \alpha_{q}=0.33$ GeV$^{-2}$.
We can use (13) for the numerical calculation of the tetraquark form factor.
It should be noted, that the calculation has not any new parameters as
compared to the calculation of the meson and tetraquark mass spectrum.
In the previous paper [29] proton charge radius was calculated.
It turned out to be $R_{p}= 0.44$ fm, that is almost a factor of two smaller than the experimental value
$R_{p\; \; \mbox {\small{exp}}}  = 0.706$ fm [32]. It is usually for the quark models with the one-gluon
input interaction [33, 34], when only the presence of the new parameters
or the introduction of an additional interaction allows to achieve a good
agreement with the experiment [35, 36].

       The behaviour of the electromagnetic form factor of the tetraquark $X(3872)$ with the mass 
$M=3872$ MeV is shown in Fig.2. The calculations were carried out for two tetraquarks with the 
masses 3872 MeV and 3940 MeV. The results turned out to be equal: the charge radius of the $X(3872)$ 
and $X(3940)$ $R_{tetra}=0.50$ fm.The charge radius was found to be approximately equal to the 
charge radius of the proton ($R_{p}= 0.44$ fm) [29]. It can be concluded that tetraquarks with charm 
are more compact systems than ordinary baryons.
\\

{\bf 4. Conclusion}\\

    The method applied in the present work based on the transition from the Feynman amplitude to the
dispersion integration over the masses of the composite particles was extended to the system of  $N$ 
quarks for the multiquark states and was applied in special case of four particle systems -- tetraquarks. 
The absence of any new parameters introduced in the model for the computation of the tetraquark form 
factors is an advantage of this method.
   \\ \\
{\bf Acknowledgment}\\

    The authors would like to thank V.I.Kochkin for useful discussions. This research was
supported by Russian Ministry of Education (Grant 2.1.1.68.26).

\newpage
{\bf References}\\ \\
$1.$\quad S. Godfrey, and S. L. Olsen, arXiv:0801.3867 [hep-ph].\\
$2.$\quad M. B. Voloshin, arXiv:0711.4556 [hep-ph].\\
$3.$\quad Belle Collaboration, S.-K. Choi {\em et al.}, Phys. Rev. Lett. {\bf 91}, 262001 

(2003).\\
$4.$\quad BaBar Collaboration, B. Aubert {\em et al.}, Phys. Rev. D {\bf 71}, 071103R 

(2005).\\
$5.$\quad CDF Collaboration, D. Acosta {\em et al.}, Phys. Rev. Lett. {\bf 93}, 072001 

(2004).\\
$6.$\quad D0 Collaboration, V. M. Abazov {\em et al.}, Phys. Rev. Lett. {\bf 93}, 162002 

(2004).\\
$7.$\quad Belle Collaboration, K. Abe {\em et al.}, Phys. Rev. Lett. {\bf 98}, 082001 

(2007).\\
$8.$\quad G. Gelmini, Nucl. Phys. B {\bf 174}, 509 (1980).\\
$9.$\quad K. T. Chao, Nucl. Phys. B {\bf 169}, 281 (1980).\\
$10.$ H. M. Chan, and H. Hogaasen, Phys. Lett. B {\bf 72}, 121 (1977).\\
$11.$ L. Maiani, F. Piccinini, A. D. Polosa, and V. Riquer, Phys. Rev. D {\bf 71}, 

014028 (2005).\\
$12.$ D. Ebert, R. N. Faustov, and O. Galkin, Phys. Lett. B {\bf 634}, 214 (2006).\\
$13.$ H. Melosh,  Phys. Rev. D {\bf 9}, 1095 (1974).\\
$14.$ G. B. West,  Ann. Phys. (N. Y.) {\bf 74}, 464 (1972).\\
$15.$ S. J. Brodsky, and G. R. Farrar,  Phys. Rev. D {\bf 11}, 1309 (1975).\\
$16.$ M. V. Terentyev, Yad. Fiz. {\bf 24}, 207 (1976).\\
$17.$ V. A. Karmanov, ZhETF {\bf 71}, 399 (1976).\\
$18.$ I. G. Aznauryan and N. L. Ter-Isaakyan, Yad. Fiz. {\bf 31}, 1680 (1980).\\
$19.$ A. Donnachie, R. R. Horgen, and P. V. Landshoft,  Z. Phys. C {\bf 10}, 71

(1981).\\
$20.$ L. L. Frankfurt, and M. I. Strikman,  Phys. Rep. C {\bf 76}, 215 (1981).\\
$21.$ L. A. Kondratyuk, and  M. I. Strikman,  Nucl. Phys. A {\bf 426}, 575 (1984).\\
$22.$ R. N. Faustov,  Ann. Phys. (N. Y.) {\bf 78}, 176 (1973).\\
$23.$ A. A. Migdal, Phys. Lett. B {\bf 7} 98 (1971).\\
$24.$ R. N. Faustov, Teor. Mat. Fiz. {\bf 3}, 240 (1970).\\
$25.$ V. V. Anisovich and A. V. Sarantsev, Yad. Fiz. {\bf 45}, 1479 (1987).\\
$26.$ S. M. Gerasyuta, Yad. Fiz. {\bf 55}, 3030 (1992). \\
$27.$ S. M. Gerasyuta, Nuovo Cimento A {\bf 106}, 37 (1993).\\
$28.$ S. M. Gerasyuta, Z. Phys. C {\bf 60}, 683 (1993).\\
$29.$ S. M. Gerasyuta, M. A. Durnev, Yad. Fiz.  {\bf 70 }, 1990 (2007).\\
$30.$ V. V. Anisovich, S. M. Gerasyuta, and A. V. Sarantsev,  Int. J. Mod. 

Phys. A {\bf 6}, 625 (1991).\\
$31.$ S. M. Gerasyuta, and V. I. Kochkin, arXiv:0804.4567 [hep-ph].\\
$32.$ M. Gourdin, Phys. Rep. C {\bf 11}, 29 (1974).\\
$33.$ A. A. Kvitsinsky {\em et. al.}, Yad. Fiz. {\bf 38}, 702 (1986).\\
$34.$ A. A. Kvitsinsky {\em et. al.}, Fiz. Elem. Chastits At. Yadra 17, 267 (1986).\\
$35.$ F. Cardarelli, E. Pace, G. Salme, and S. Simula, Phys. Lett. B {\bf 357}, 267 

(1995).\\
$36.$ F. Cardarelli, E. Pace, G. Salme, and S. Simula, nucl-th/9809091.\\

 \newpage
\begin{center}
 \includegraphics[scale=0.6]{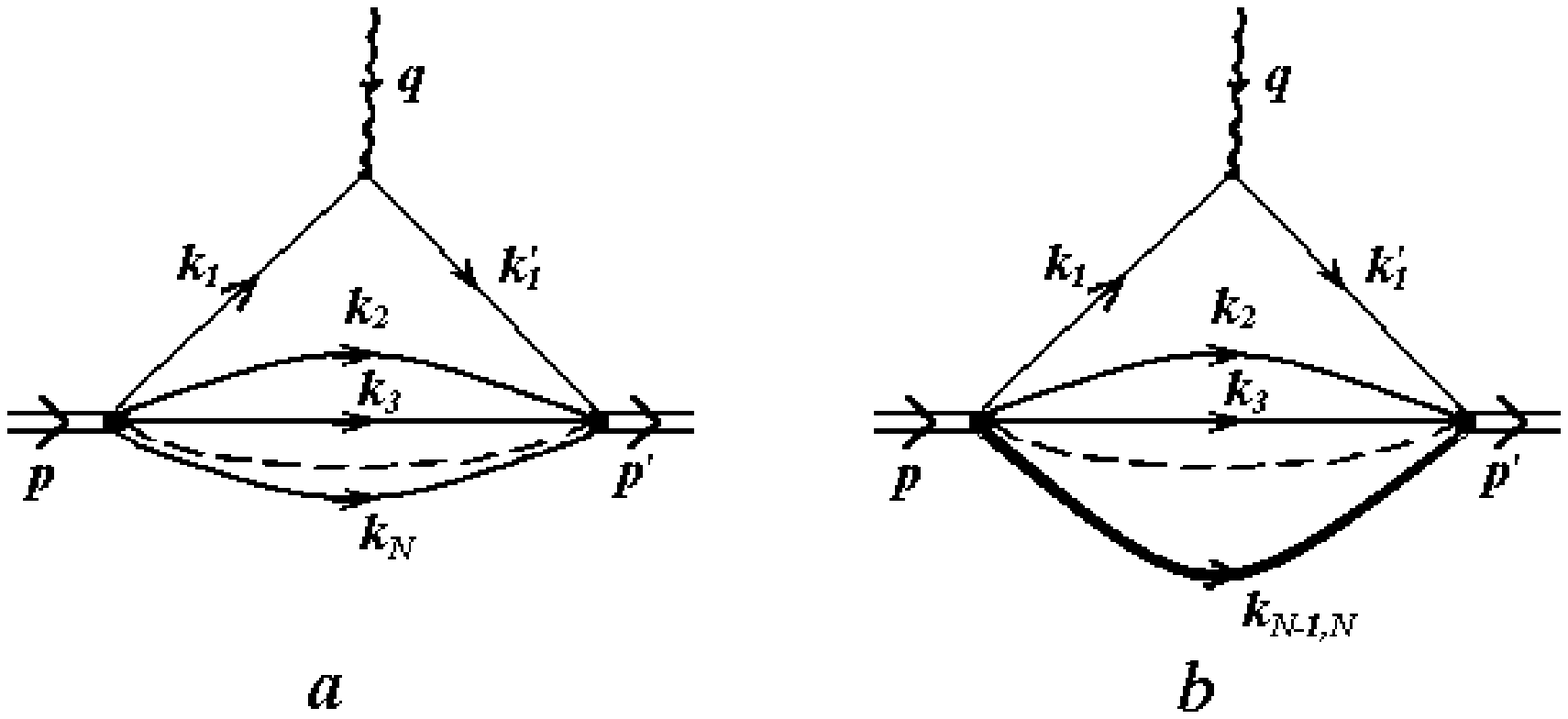}
 \end{center}
 \newpage
\begin{center}
 \includegraphics[bb=0 600 300 250,scale=0.5]{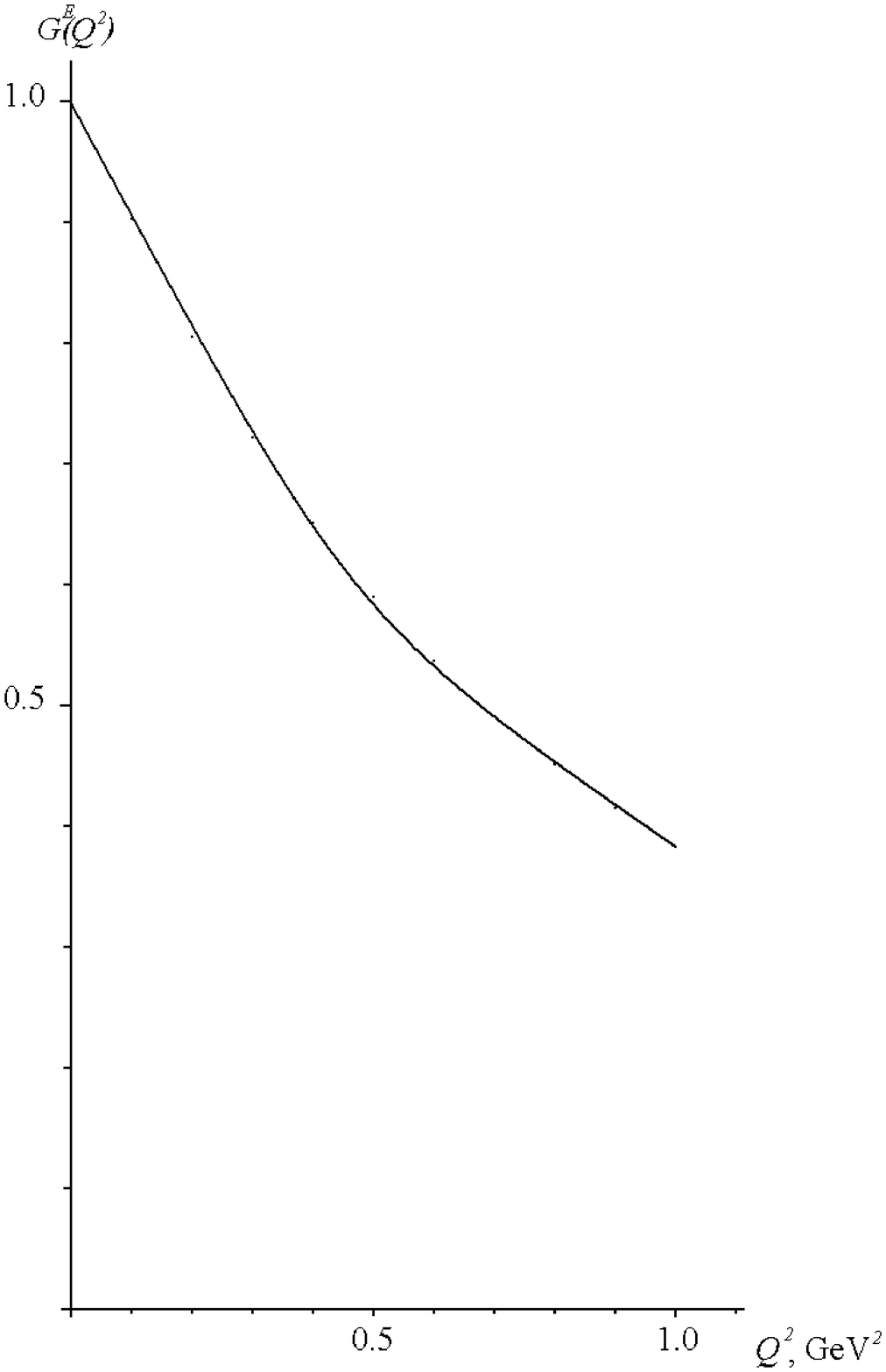}
\end{center}
\newpage

{\bf Figure captures} \\ \\

{\bf Fig.1} Triangle diagrams, which determine the form factors of tetraquarks.\\

{\bf  Fig.2} The electromagnetic form factor of the tetraquark $X(3872)$  with 

mass M=3872 MeV.

\end{document}